\documentclass[useAMS,usenatbib,usegraphicx]{mn2e}

\usepackage{amsmath, amssymb, graphics, graphicx, rotating, lscape}

\title[CP spectrum of PKS\,B2126--158]
{Broadband radio circular polarization spectrum of the relativistic jet in PKS\,B2126--158}

\author[O'Sullivan et al.]{S.~P.~O'Sullivan$^{1}$, N.~M.~McClure-Griffiths$^{2}$, I.~J.~Feain$^{1,2}$, B.~M.~Gaensler$^{1}$, R.~J.~Sault$^{3,4}$ \\ \\
$^{1}$Sydney Institute for Astronomy, School of Physics, The University of Sydney, NSW 2006, Australia.\\
$^{2}$CSIRO Astronomy and Space Science, ATNF, PO Box 76, Epping, NSW 1710, Australia. \\
$^{3}$School of Physics, The University of Melbourne, Parkville, VIC 3010, Australia. \\
$^{4}$National Radio Astronomy Observatory, P.O. Box O, Socorro, NM 87801, USA. \\
}
\begin{document}

\date{Accepted 2013 July 14.  Received 2013 July 12; in original form 2013 March 27}
\pagerange{\pageref{firstpage}--\pageref{lastpage}} \pubyear{2013}
\maketitle
\label{firstpage}

\begin{abstract}

We present full-Stokes radio polarization observations of the quasar
PKS\,B2126--158 ($z=3.268$) from 1 to 10~GHz using the Australia Telescope Compact Array. 
The source has large fractional circular polarization, $m_c\equiv|V|/I$,   
detected at high significance across the entire band (from 15 to 90$\sigma$ per 128~MHz sub-band). 
This allows us to construct the most robust circular polarization (CP) spectrum of an AGN jet to date. 
We find $m_c\propto\nu^{+0.60\pm0.03}$ from 1.5 to 6.5~GHz, with a peak 
of $m_c\sim1\%$ before the spectrum turns over somewhere between 
6.5 and 8~GHz, above which $m_c\propto\nu^{-3.0\pm0.4}$. 
The fractional linear polarization ($p$) varies from $\lesssim0.2\%$ to $\sim$1\% across 
our frequency range and is strongly anti-correlated with the fractional CP, with 
a best-fit power law giving $m_c\propto p^{-0.24\pm0.03}$. 
This is the first clear relation between the observed linear and circular polarisations 
of an AGN jet, revealing the action of Faraday conversion of linear polarization (LP) 
to CP within the jet. 
More detailed modelling in conjunction with high-spatial resolution observations are required to determine 
the true driving force behind the conversion (i.e. magnetic twist or internal Faraday rotation). 
In particular determining whether the observed Faraday rotation is internal or entirely external to the jet is key to this goal. 
The simplest interpretation of our observations favours some internal Faraday rotation, implying that Faraday rotation-driven 
conversion of LP to CP is the dominant CP generation mechanism. In this case, a small amount of vector-ordered 
magnetic field along the jet axis is required, along with internal Faraday 
rotation from the low energy end of the relativistic electron energy spectrum in an electron-proton 
dominated jet. 

\end{abstract}

\section{Introduction}

Large-scale, ordered magnetic fields are invoked to explain the launching, acceleration 
and collimation of relativistic jets from the central nuclear region of active galactic nuclei (AGN) 
\citep{meierjapan}. These jets of relativistic plasma can be formed by strong magnetic fields 
centrifugally lifting particles out of the accretion disk \citep{blandfordpayne1982} and/or accelerating a pair 
plasma cascade near the black-hole horizon \citep{blandfordznajek1977}. However, the 
three-dimensional jet magnetic field structure and particle composition are still not well 
constrained observationally. Circular polarization (CP), measured as Stokes $V$, in the radio continuum 
emission from AGN jets is a powerful diagnostic of the jet magnetic field and 
particles since, unlike linear polarization, it is expected to remain almost completely 
unmodified by external screens 
\citep[although see][for the special case of scintillation induced CP]{macquartmelrose2000}.  
 
The integrated emission of radio-loud AGN typically show linear polarizations of a 
few percent of the total intensity (Stokes $I$), 
while detections of the degree of CP generally find values ranging from 
0.1--0.5\% \citep{saikiasalter1988}, with values as large as 1\% being uncommon \citep{aller2012}. 
From a compilation of reliable measurements by \cite{weilerdepater1983} and the monitoring 
observations of \cite{komesaroff1984}, we know that CP fractions are 
higher in flat spectrum sources, the CP is more variable than both total intensity and 
linear polarization (LP) fraction, there is no clear correlation between LP and CP, and there is a preferred 
handedness in individual sources. 
\cite{rayner2000} confirmed these results with much higher accuracies and also found 
that the CP spectral index between 1.4 and 4.8~GHz was approximately flat ($\alpha_{m_c}=0.1\pm0.3$, 
where the CP spectral index ($\alpha_{m_c}$) is defined as $m_c\propto\nu^{\rm{+}\alpha_{m_c}}$, with 
$m_c\equiv |V|/I$ defining the degree of CP) for a sample of 12 AGN. 
Substantial progress has also been made through high spatial resolution CP measurements 
using VLBI \citep{homanwardle1999, homan2001,homanlister2006, vitrishchak2008}. 
It was found that local levels of CP could be quite strong on VLBI scales, with detected 
CP fractions generally ranging from 0.3--1\%. Most detections were found coincident with the VLBI core, 
presumably indicating that the CP generation mechanism is most efficient near the unity optical depth surface. 
\cite{vitrishchak2008} quoted CP spectral indices from measurement at 15, 22 \& 43~GHz 
but were not able to use them to find any clear trend pointing to a specific CP generation mechanism. 

Our ability to extract physical insights about the nature of jets from CP observations depends 
on our ability to determine the dominant mechanism for CP production. 
There are many predictions for the origin of the CP emission in radio-loud AGN. 
Synchrotron emission itself has an intrinsic level of CP \citep{leggwestfold1968}, however, 
a separate generation mechanism in which Faraday conversion of LP to CP 
occurs, can potentially dominate the observed emission \citep{jonesodell1977b, jones1988}. 
There are several mechanisms through which this Faraday conversion can occur in the jet: 
Faraday rotation internal to the jet due to either thermal electrons or the low energy end of the 
relativistic electron energy spectrum \citep[e.g.,][]{pacho1973,jonesodell1977b}, 
a change in the orientation of the perpendicular component of the magnetic 
field along the line of sight \citep{hodge1982,ensslin2003,gabuzdacp2008}, or anisotropic 
turbulence in the jet where the anisotropy is created from a net poloidal magnetic flux 
\citep{ruszkowski2002,beckertfalcke2002}. 
\cite{homan2009} studied the 
full-polarization spectra of the quasar 3C~279 at 6 frequencies between 8 and 22~GHz and used an 
inhomogeneous conical jet model \citep{blandfordkonigl1979} dominated by a vector-ordered poloidal 
field to estimate the net magnetic flux carried by the jet, as well as the jet particle content and lower 
cutoff in the relativistic electron energy spectrum. 
Furthermore, through CP measurements it may be possible to get observational proof of whether the 
magnetic field structure has its roots in the magnetic field of the supermassive 
black-hole/accretion-disk system responsible for giving rise to the jets \citep{ensslin2003}. 
For detailed reviews of both theory and observation of CP in AGN, see \cite{macquart2002} and 
\cite{macquartfender2004}.

The Australia Telescope Compact Array (ATCA), with its recent Compact Array Broadband 
Backend (CABB) system and centimetre receiver upgrades, allows high precision polarimetry 
over wide frequency bandwidths, making it uniquely placed to answer fundamental questions 
about the nature of circular polarized emission from AGN jets. 
The focus of this paper is on 1 to 10~GHz observations with the ATCA of the GigaHertz-Peaked Spectrum (GPS) quasar, 
PKS\,B2126--158 \citep{stanghellini1998}. 
PKS\,B2126--158 (J2129-1538) is one of the brightest quasars known ($m_V=17.1$), 
with a redshift of $z=3.268$ \citep{jauncey1978,osmer1994}, allowing detailed optical 
studies of absorption systems along the line of sight \citep{dodorico1998}. 
In a recent study of its broadband spectral energy distribution, \cite{ghisellini2010} 
estimated a very large black hole mass of $10^{10}$~M$_\odot$. 
\cite{raynerphd} found this source to be strongly 
circularly polarized, with $V/I=-1.403\pm0.008\%$ at 4.8~GHz, using the old 
ATCA backend system (128~MHz instantaneous bandwidth). 
Hence, we specifically targeted PKS\,B2126--158 to investigate the frequency dependence 
and origin of its CP emission.

\section{Observations and Data Reduction}

PKS\,B2126--158 was observed with the standard ATCA continuum mode \citep{cabbpaper} 
over the frequency ranges 1 to 3~GHz, 4.5 to 6.5~GHz and 8 to 10~GHz in the 6A array configuration over $\sim$24 hrs 
from 2011 April 30 to May 1. Two other target sources were also observed, the results of which will be 
published elsewhere. The total integration time for PKS~B2126-158 in the 1 to 3~GHz band was 215~mins 
while the source was observed for $\sim$130~mins simultaneously at 4.5 to 6.5~GHz and 8 to 10~GHz. 
Our resolution varies from $\sim$$10''$ (1~GHz) to $1''$ (10~GHz) with the source 
remaining spatially unresolved on all baselines and at all frequencies. 

For the linear feed system on the ATCA, it is mainly the instrumental polarization 
leakages that cause corruption of Stokes $V$ by Stokes $I$, while 
gain errors lead to leakage of Stokes $Q$ and $U$ into $V$. 
Therefore, instrumental leakages need to be determined with high precision 
for the ATCA to obtain accurate circular polarization measurements. 
A positive/negative value of Stokes $V$ corresponds to a right-handed/left-handed circularly 
polarized wave (IAU Transactions, 1974, XVB). 
We define the Stokes $I$ spectral index, $\alpha$, such that the observed flux 
density ($I$) at frequency $\nu$ follows the relation $I_{\nu}\propto\nu^{\rm{+}\alpha}$. 
Calibration was done using the \textsc{Miriad} data reduction package \citep{sault1995} 
in a similar fashion at all three frequency bands. The bandpass and flux 
calibration was achieved using the primary calibrator PKS~B1934-638 in 
all cases. The \textsc{Miriad} task \textsc{Mirflag} was used for automated 
flagging in conjunction with some minor manual flagging. 
To calibrate the polarization we followed the procedure outlined by 
\cite{rayner2000} and \cite{raynerphd}, adjusted for the new wide 
bandwidth system as described below.  

From 1 to 3~GHz, we used PKS~B2326-477 as the leakage calibrator with seven several-minute 
cuts spread over the observation period, which provided $\sim$$90^\circ$ of parallactic 
angle coverage. 
To obtain complete calibration solutions for all polarization leakage terms requires not only 
large parallactic angle
for the leakage calibrator but also needs it to be strongly polarized ($>3\%$). 
PKS~B2326-477 remains polarized at $\sim$5\% across the entire 
1 to 3~GHz band making it an excellent leakage calibrator for this band. 
The main difference from \cite{rayner2000} is that the instantaneous bandwidth 
of the ATCA is now 16 times larger. Therefore, we obtained 
independent calibration solutions for 16 sub-bands of 128~MHz to determine 
the frequency-dependent leakages, 
and to correct for the gain-variation and spectral slope effects across the 2~GHz bandwidth.  
The 1 to 3~GHz band is strongly affected by RFI from $\sim$1.1 to 1.4~GHz so we 
have avoided analysing this frequency range due to the substantial amount of 
data lost to flagging. 
The observations from 4.5 to 6.5~GHz and 8 to 10~GHz were conducted simultaneously 
and the leakage calibrator used for this band was PKS~B2005-489. Excellent 
parallactic angle coverage ($\sim$$160^\circ$) was obtained for this calibrator 
and its percentage linear polarization ranges from $\sim$3\% to 5.5\% between 4.5 and 10~GHz. 

After the calibration solutions were copied to the target, PKS\,B2126--158, we generated 
Stokes $I$, $Q$, $U$, and $V$ clean images at 128~MHz intervals, from which we extracted 
the peak flux. A small region around 
the position of the source in the clean-residual map was used to provide an estimate of the 
contributions to the error from both the system noise and any residual calibration 
artifacts. 
The linear fractional polarization was constructed as $p=((Q/I)^2+(U/I)^2)^{1/2}$ and the linear polarization 
angle as $\Psi=\frac{1}{2}\tan^{-1}(U/Q)$. 
The typical error measured in the 128~MHz images for both Stokes $V$ ($\sigma_V$) and 
the linear polarized flux ($\sigma_p$) was approximately 
0.1, 0.2, 0.2~mJy~beam$^{-1}$ in the 1 to 3, 4.5 to 6.5 and 8 to 10~GHz bands, respectively 
($\sim$3 times the theoretical rms noise level). 
The linear polarized flux was corrected for Ricean 
bias \citep{wardlekronberg1974}. Data points with fractional linear polarisation less than $5\sigma$ were 
discarded (this only affected some of the linear polarization data in the 4.5 to 6.5 GHz range).

Similar to flux calibration, the CP zero-level must be determined from 
a source with known CP (e.g., exactly 0\%, or some 
other value known to within the level of precision required). 
The zero-point CP has been set assuming that PKS~B1934-638 has 
Stokes $V=0$, which is the usual approach in \textsc{Miriad}. 
\cite{rayner2000}, however, claimed that PKS~B1934-638 is circularly polarized, 
albeit at the relatively small level of approximately $+0.03\%$ at both 1.4 and 4.8~GHz. 
This means that all CP observations are biased by a fractional CP of 
equal magnitude and opposite sign to that of PKS~B1934-638; although 
since CP is highly variable this value may have changed since. 
For our present analysis, we are not overly concerned about potential contamination 
from this small level of CP, although we do include the CP zero-level of 
PKS~B1934-638 as an additional error in the derived values of $m_c$, 
where the error in $m_c$ is defined as $\sigma_{m_c}^2=(\sigma_V/I)^2 + m_{c,{\rm 1934}}^2$, 
where $m_{c,{\rm 1934}}=3\times10^{-4}$. 


\begin{figure}
\centering
\vspace{-0.5cm}
    \includegraphics[width=8.5cm]{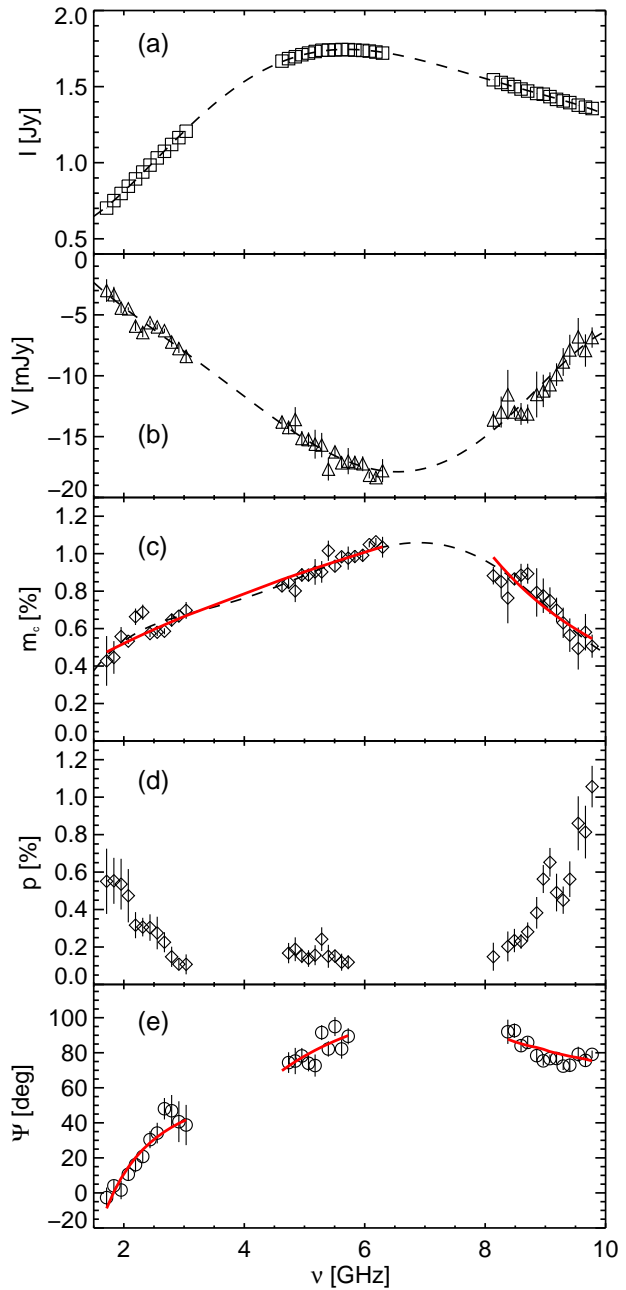}
      \caption{Integrated spectrum of PKS\,B2126--158 from 1.5 to 10 GHz with data points 
      plotted in 128~MHz intervals: 
      (a) Stokes $I$ in Jy, with $3\sigma_I$ error bars (too small to be visible) 
      (b) Stokes $V$ in mJy, with $3\sigma_V$ error bars 
      (c) degree of circular polarization $m_c\equiv|V|/I$ in per cent, with $3\sigma_{m_c}$ error bars
      (d) degree of linear polarization, $p$, in per cent, with $3\sigma_p$ error bars after correction for Ricean bias 
      ($p<5\sigma$ data not shown)
      (e) linear polarization angle ($\Psi$) in degrees, with $3\sigma_{\Psi}$ errors bars 
      ($\Psi$ data values not shown where $p<5\sigma$). 
      See Section~2 for definitions of errors. 
      The dashed lines correspond to fifth-order polynomial fits to the data. 
      The solid lines (red) in panel (c) represent power-law fits to the data (Section 3), while in panel (e) 
      the solid lines (red) represent the best-fit Faraday rotation law (see Section~3.2 and Figure~4 for details). 
}
  \label{2126}
\end{figure}

\section{Results \& Discussion}

Figure~\ref{2126} shows the full Stokes properties of PKS\,B2126--158 from 1.5 to 10~GHz. 
The total intensity spectrum is shown in Figure~\ref{2126}a, with the dashed line 
representing a fifth-order polynomial fit to the data. The inverted part of the spectrum from 1 to 3~GHz 
has a spectral index of $\alpha=+0.945\pm0.001$ (by fitting a power-law to the data in this range). The spectrum 
turns over at $\nu_t=5.7\pm0.1$~GHz with a peak flux of $I_{max}=1.744\pm0.001$~Jy 
(measured from a parabolic fit to the 4.5 to 6.5~GHz data only). 
A power-law fit to the steep part of the spectrum at the high frequency band of 8 to 10~GHz 
found $\alpha=-0.708\pm0.003$.  
The observed levels of Stokes $V$ display a smooth variation across the entire band, 
with a peak value of $\sim-18$~mJy ($90\sigma$) at 6.5~GHz and turning over somewhere 
between 6.5 and 8~GHz (Figure~\ref{2126}b). Even the lowest detected level of CP flux 
of $\sim3$~mJy is at high significance ($15\sigma$). The sign of Stokes $V$ corresponds 
to left-circularly polarized emission and does not change across the band. 
For $\nu<6.5$~GHz, we fit a power-law to the percentage CP ($m_c$), finding a 
CP spectral index $\alpha_{m_c}=+0.60\pm0.03$. For $\nu>8$~GHz, we find 
$\alpha_{m_c}=-3.0\pm0.4$. Both fits are represented by the solid lines on 
Figure~\ref{2126}c. We consider these fits as the most robust measurements 
of the CP spectrum of an extragalactic source to date. 
The high fractional CP of 0.5 to 1\% seen here is uncommon in AGN, with values approaching 1\% and greater 
being very rare \citep{macquart2000, homanwardle2004, vitrishchak2008}. 
Previous measurements of large CP from the intra-day variable source PKS B1519-273 \citep{macquart2000}, 
determined a CP spectral index $\alpha_{m_c}=0.7_{-0.3}^{+1.4}$ from observations 
at 1.4, 2.3, 4.8 and 8.4~GHz. Since this is consistent with our observed CP spectrum, 
below the turnover, it potentially indicates a common generation mechanism in both these sources. 
 
\begin{figure}
\centering
    \includegraphics[width=8.0cm]{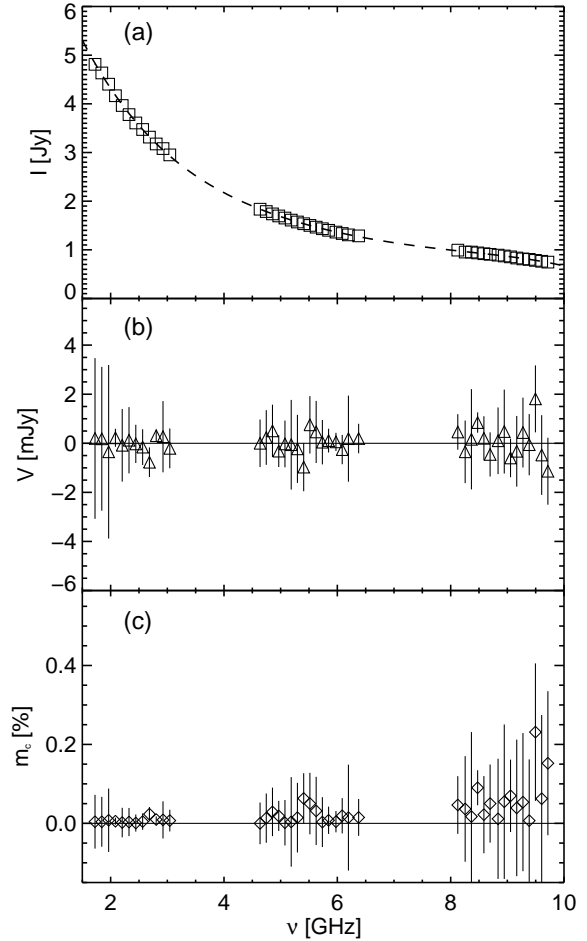}
      \caption{Integrated spectrum of PKS~B0252-712 from 1.5 to 10 GHz with data points 
      plotted in 128~MHz intervals: 
      (a) Stokes $I$ in Jy, (b) Stokes $V$ in mJy, (c) degree of circular polarization ($m_c$), in per cent. 
      The error bars are plotted in the same manner as in Figure~1.
      No significant levels of Stokes $V$ are detected above $5\sigma_V$ for this source. 
      Plots of linear polarization are not included as there were no detections above $5\sigma_p$. 
      The dashed line in panel (a) corresponds to a fifth-order polynomial fit to the data. 
      The solid lines in panels (b) and (c) are not fits to the data but simply drawn at zero. 
}
  \label{0252}
\end{figure}

As a check of the reliability of our results we also show the derived spectrum for the Compact, Steep-Spectrum 
(CSS) source PKS~B0252-712 \citep{tzioumis2002}. This source was observed contemporaneously and 
exactly the same calibration solutions were applied as for PKS\,B2126--158 in each band.  
As is clearly seen in Figure~\ref{0252}, no significant levels of CP are detected. 
This is not unexpected given that its emission is dominated by 
a compact, double-lobed structure \citep[][fig.~5]{tzioumis2002}. 
It is also worth noting that, considering the relatively low fractional LP levels of PKS\,B2126--158, 
it is not conceivable that the measured Stokes $V$ is leakage from Stokes $Q$ and $U$.  
This gives us full confidence in the reliability of our CP measurements for 
PKS\,B2126--158 as well as its broadband frequency dependence. 
We note that the classification of this source as a GPS quasar \citep{stanghellini1998} 
has the potential for confusion with 
GPS galaxies, from which such high CP would be very surprising given that Doppler boosting 
is not expected to play a significant role. 
GPS galaxies are an homogenous class of sources consisting mainly of compact 
symmetric objects in which the integrated emission is dominated by their extended regions \citep{stanghellini2003}. 
GPS quasars, on the other hand, can be considered to be intrinsically similar to flat-spectrum radio 
quasars (FSRQs) that show a core-jet type morphology on milliarcsecond scales. 
In this case, a small number of components, close 
to the base of the jet, dominate the radio emission and lead to a turnover in the integrated 
spectrum due to either synchrotron self-absorption \citep{mutel1985} or free-free absorption 
\citep{bicknell1997}.

\subsection{Spatial location of the emission}

PKS\,B2126--158 has been observed at high angular resolution ($\sim$5~mas) with VLBI 
\citep{fomalont, scott, charlot}, showing that the total flux is almost 
completely dominated by the core with only a small extension of emission to the south. 
This indicates a potentially very small angle to the line of sight for this source and hence, 
strong Doppler boosting of the approaching jet emission. 
\cite{raynerphd} found an integrated flux of $\sim$1.2~Jy at 4.8~GHz with the ATCA in Oct 1995, 
while monitoring of this source by \cite{tingay} with the ATCA between 1996 and 2000 found approximately 
the same flux, with the source exhibiting no significant variability during this time. 
\cite{fomalont} also found a total flux of $\sim$1.2~Jy from model-fitting 2-D Gaussian 
components to VLBA observations at 5~GHz in June 1996. 
From this we can conclude that there is no significant flux on intermediate angular scales for this 
source and practically all the emission we observe with the ATCA is coming from 
the compact inner jet regions (i.e. within $\sim$5~mas, which corresponds to a projected 
linear size of $\sim$$40$~pc). Therefore, for this particular source, both the Stokes $I$ and 
Stokes $V$ emission we detect is most likely coming from the bright, unresolved VLBI core. 
This is consistent with the vast majority of VLBI CP observations \citep{homanlister2006, vitrishchak2008} 
and, importantly, gives us confidence that our CP spectral indices would be the same 
as those derived if the observations were conducted with the high spatial resolution of VLBI. 
It is important to note that this source appears to have flared around 2009\footnote{http://www.narrabri.atnf.csiro.au/calibrators/} 
with its 5~GHz Stokes $I$ flux increasing to its current level of $\sim$1.7~Jy. 
However, the amount of Stokes $V$ appears to have changed very little, since 
the decrease in $m_c$ from $\sim$1.4\% \citep{raynerphd} to $\sim$0.9\% at 4.8~GHz 
can be explained almost purely by the increase in the Stokes $I$ flux alone.

\begin{figure}
\centering
    \includegraphics[width=8.5cm]{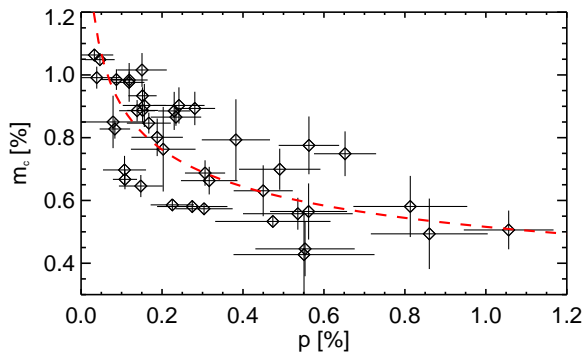}
      \caption{Plot of degree of linear polarization ($p$) versus the 
      degree of circular polarization ($m_c$) at each frequency measurement. 
      A Spearman rank correlation of $-0.8$ quantifies the strong anti-correlation 
      between the two variables.  
      The dashed line (red) represents the best-fit power-law with $m_c\propto p^{-0.24\pm0.03}$. 
}
  \label{correlation}
\end{figure}

\begin{figure}
\centering
    \includegraphics[width=8.5cm]{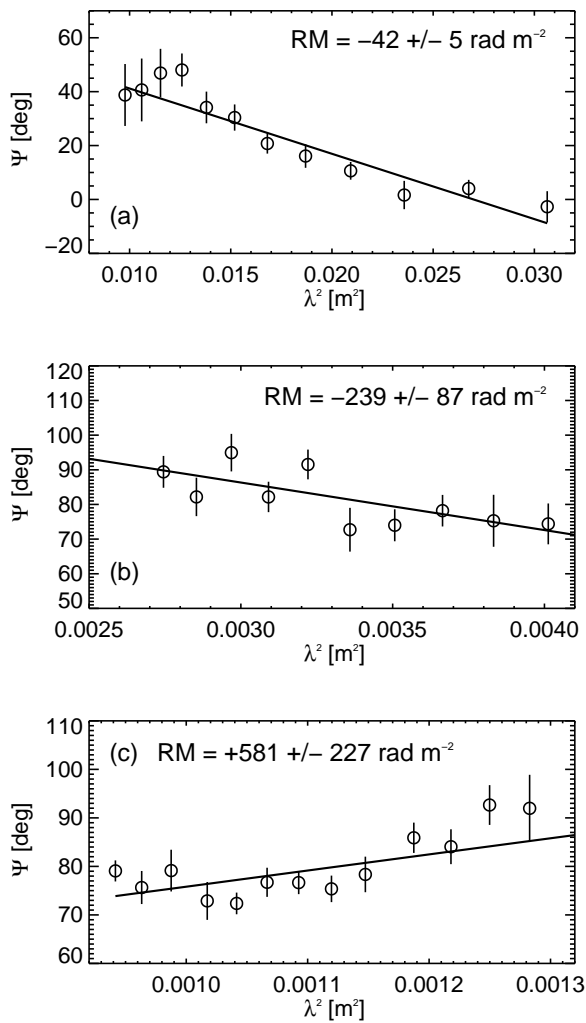}
      \caption{Plots of linear polarization angle ($\Psi$) vs.~wavelength squared ($\lambda^2$) 
      in three frequency bands, with the solid lines in each panel representing the best-fit 
      linear $\Psi(\lambda^2)$-law describing the Faraday rotation measure (RM) in each band: 
      (a) 1.5 to 3 GHz
      (b) 4.5 to 6.5 GHz 
      (c) 8 to 10 GHz.
}
  \label{RMplots}
\end{figure}

\subsection{Consideration of the Linearly Polarized Emission}

Figure~\ref{2126}d,e shows the frequency dependence of the linear polarized 
emission over the same range as the total intensity and circular polarization. 
We note the low degree of linear polarization ($p\lesssim0.2\%$) from 4.5 to 6.5~GHz 
and the surprisingly steep rise toward $p\sim1\%$ at both ends of the band, as well 
the non-monotonic distribution of linear polarization angles ($\Psi$) with frequency.  
Assuming that the linear polarized emission we detect is coming from the same 
region of the jet as the CP, then the ratio of $m_c$ to $p$ also reaches surprisingly large 
values, ranging from $\sim$0.5 to $\sim$10 across our full frequency coverage. 

The inverted spectrum of the Stokes $I$ emission from 1.5 to 6.5~GHz (Figure~1a) 
indicates that the observed emission in these frequency bands originates from a 
large range of optical depths in the jet. If we consider the most commonly used jet model of \cite{blandfordkonigl1979}, 
then the position of the unity optical depth surface is frequency dependent (i.e.~$r_{\tau=1}\propto\nu^{-1}$). 
Hence, any analysis of the polarized emission from 1.5 to 6.5~GHz is complicated 
by the fact that the observed emission at different frequencies is likely coming from 
different regions in the jet. Numerical radiative transfer modelling is required to 
properly analyse the polarised emission from these regions and we defer such 
detailed analysis to future work. For the remainder of this paper we conduct 
a more qualitative analysis, where we will refer to the ``optically thick'' regime as corresponding to 
emission at $\nu<6.5$~GHz and the ``optically thin'' regime, where the spatial 
location of the emission does not change with frequency, corresponding to $\nu>8$~GHz. 

In Figure~\ref{2126}c,d the percentage linear 
polarization ($p$) appears anti-correlated with the percentage CP ($m_c$), 
so in Figure~3, we plot $p$ versus $m_c$ for each frequency measurement. 
A Spearman rank correlation coefficient of $-0.8$ quantifies the strong 
anti-correlation between these two variables, where a 
correlation coefficient of $-1$ occurs in the case of a perfect anti-correlation. 
This is the first time such a relation between the LP and CP of an AGN jet has 
been observed, and strongly supports the action of Faraday conversion 
of LP to CP. Fitting a power-law dependence to the data gives a best-fit 
relation $m_c\propto p^{-0.24\pm0.03}$ (Fig.~3). Such a dependence has not 
been predicted in previous theoretical work. However, the degree of LP can 
also be strongly affected by the frequency dependent effects of Faraday depolarisation 
as well as optical depth effects, which makes this relation non-trivial to analyse. 
Separate power-law fits to the $m_c$ versus $p$ data in the optically-thick  and optically-thin 
regimes gives $m_c^{thick}\propto p^{-0.33\pm0.04}$ and 
$m_c^{thin}\propto p^{-0.23\pm0.05}$, respectively. This may lead one to 
naively suggest that conversion is more efficient in the optically-thin regime; however, 
it is more likely that different linear polarisation characteristics due to optical depth 
effects are responsible. 
In any case, the scatter in the data is large and it is premature to draw any definite conclusions 
about differences between conversion efficiency in the optically thick/thin regimes. 

For both the intrinsic CP and Faraday conversion of LP to CP mechanisms, 
the large CP excess (ratio of CP to LP), noted at the beginning of this section, 
is not generally expected \citep[e.g.][]{jonesodell1977b,valtaoja1984}. 
The low observed levels of fractional linear polarization are most likely 
explained by either strong depolarization from internal Faraday rotation and/or large 
rotation measure gradients in an inhomogeneous external screen. 
In general, the distribution of Faraday rotating material in the immediate vicinity of AGN jets is 
not very well constrained. Most studies conclude that the majority of rotating 
material is contained in a sheath of slower moving, thermal material surrounding the jet 
\citep[e.g.][and references therein]{osullivangabuzda2009, broderickloeb2009, porth2011}. 
Although a significant contribution from unrelated ionized gas clouds, that clearly exist in some 
sources \citep{walker2000}, cannot be discounted. 
Large amounts of Faraday rotation from thermal plasma within jets is generally considered 
unlikely on energetic grounds \citep{celotti1998}, and we also consider it unlikely for this 
particular source (see Section 3.3 for details).

Recent work by \cite{farnsworth2011} and \cite{osullivan2012} shows that fitting various Faraday rotation models 
simultaneously to the $Q/I$ and $U/I$ data provides the most robust estimation of RM. However, this 
implicitly assumes a negligible contribution from conversion of LP to CP. Since the frequency dependence of 
our Stokes $Q$ and $U$ data is clearly contaminated by the large amount of Stokes $V$ in this source, we 
revert to the traditional way of finding the RM, by fitting a line to the variation in the linear polarization angle with wavelength 
squared ($\lambda^2$), where the change in the linear polarization angle, $\Delta\Psi = {\rm RM}\lambda^2$. 
No single RM can fit all the data from 1 to 10~GHz. Instead we determine an RM for each the 
observing bands individually, finding RMs of $-42\pm5$~rad~m$^{-2}$, $-239\pm87$~rad~m$^{-2}$ and 
$+581\pm227$~rad~m$^{-2}$ from 1.5 to 3~GHz, 4.5 to 6.5~GHz and 8 to 10~GHz, respectively 
(fits shown in Figure~4 and also represented in Figure~1e). The 8 to 10~GHz data is not very well 
described by a linear $\Psi(\lambda^2)$-law and further observations at shorter wavelengths are required 
to test the presence of any complex behaviour. 
The change in sign of the fitted RM in the 8 to 10~GHz band is not too difficult to explain since the sign  
simply depends on whether the dominant line-of-sight component of the magnetic field is 
pointing towards or away from the observer, which could potentially change 
as we sample the full radiation depth of the jet and/or as we probe further upstream in the jet 
at these higher frequencies. Considering just the magnitude of the RM it is clear that it increases 
systematically towards higher frequencies and can be fit by a power-law, with $|{\rm RM(\nu)}|\propto \nu^{+1.9\pm0.1}$. 

Frequency-dependent RMs have previously been measured in jets whose observed polarization is dominated 
by emission from the spatially-unresolved VLBI core at the base of the jet \citep{jorstad2005, osullivangabuzda2009, algaba2011} 
and through synthetic observations in jet simulations of such regions by \cite{porth2011}. 
Such an effect is expected \citep{wardlehoman2003} when the electron number density and magnetic 
field scale in such a way along (or outside) the jet as to cause a constant rotation of the linear polarization angle 
(in this case $\sim40^\circ$). 
To illustrate this effect, we consider power-law scalings of the magnetic field ($B$)
and electron number density ($n_e$) with distance ($r$) from the base of the jet. 
In a conical jet, we expect $B\propto r^{-1}$ and $n_e\propto r^{-2}$, hence, we get 
${\rm RM}\propto n_e B\,r \propto r^{-2}$. As noted already, the observed jet emission is 
dominated by emission from the unity optical depth surface in a conical jet, where $r_{\tau=1}\propto\nu^{-1}$. 
Therefore, we expect the RM to scale as $\nu^{+2}$ and $\Delta\Psi\sim{\rm constant}$, as 
observed. 
This has important implications for our analysis as it means that the particles and magnetic fields 
dominating the observed Faraday rotation are most likely internal to the jet. However, we cannot 
rule out that a boundary layer surrounding the jet scales in a similar fashion as is expected within 
the jet. 
We note that the source is located at a relatively high Galactic latitude ($-42^\circ$) where the 
Galactic RM contribution can be ignored (estimated at $<10$~rad~m$^{-2}$ using \cite{taylor2009} in this 
region of the sky).

\subsection{Circular polarization generation mechanism}

Through our measurements, there are several fundamental aspects relating to the 
CP generation mechanism on which we can comment. 
Firstly, we can conclusively rule out intrinsic CP from a homogeneous source region 
as the dominant source of CP, since the spectrum does not follow the expected frequency 
dependence of $m_c\propto \nu^{-1/2}$ in the optically-thin regime \citep{leggwestfold1968} and Stokes $V$ 
does not change sign near the Stokes $I$ turnover frequency \citep{melrose1972}. 
For conversion of LP to CP in a homogeneous 
source, \cite{kennettmelrose1998} predict $m_c\propto \nu^{-1}$ for a co-spatial 
relativistic and cold plasma or $m_c\propto \nu^{-3}$ for purely relativistic plasmas. 
In the optically thin regime (i.e. $\nu>8$~GHz), we find $m_c\propto \nu^{-3.0\pm0.4}$. 
This strongly supports conversion of LP to CP by the relativistic particles and 
magnetic field within the jet as the dominant CP generation mechanism in this source 
(and rules out any significant contribution from thermal plasma within the jet). 
However, we cannot determine the exact jet composition from this model, since such a 
dependence of $m_c(\nu)$ can occur in a normal plasma (relativistic electrons 
and protons) or in a pure pair-plasma (relativistic electrons and postitrons).\footnote{AGN jets are 
considered to be effectively charge-neutral, with roughly equal numbers of relativistic 
electrons and protons (or positrons) flowing outwards \citep{begelmanblandfordrees1984}.}
 
The frequency dependence of $m_c$ below the turnover (from 1.5 to 6.5~GHz) is more difficult to 
analyse than the dependence above the turnover, because of the need for polarized radiative transfer simulations to 
accurately model the physical properties of the jet. 
Such simulations were done by \cite{jonesodell1977b} and \cite{jones1988}, 
predicting that both conversion and synchrotron emission can contribute to 
the observed CP. In the case of inhomogeneous source regions, they found that 
conversion was the dominant process, with the CP strongest near the optically 
thick core at the base of the jet. 
The fact that our observed $m_c$ peaks near the frequency at which the 
Stokes $I$ emission turns over is a strong indication that conversion is the 
dominant generation mechanism for this source. 
Furthermore, we do not observe a $90^\circ$ LP angle flip, when going from 
the optically thin to optically thick regime across the 
turnover.\footnote{The naive expectation of a $90^\circ$ change in the LP 
angle can be understood qualitatively by resolving the electric field vector into 
components parallel and perpendicular to the magnetic field. 
In the optically thin regime, the linear polarization vector is mostly in the plane 
of gyration of the electrons (perpendicular to the magnetic field). 
However, in the optically thick regime, the component of the linear polarization 
perpendicular to the magnetic field has a much higher probability of being absorbed (with respect to the 
parallel component). Therefore, in the optically thick regime, the observed 
plane of polarization is parallel to the magnetic field, leading to a $90^\circ$ 
change in the observed linear polarization angle when transitioning between the two regimes. }
This is also consistent with the \cite{jonesodell1977b} results, which show 
that simple boundary effects can eliminate both the $90^\circ$ change of the LP 
position angle and the sign flip for CP.
This links both the LP and CP emission 
regions to the brightest part of the jet near the unity optical depth surface.  
 
In a normal relativistic plasma, conversion is caused by Faraday rotation within the jet, 
where in this case, the Faraday rotation would be dominated by the contribution 
from the low energy end of the relativistic electron energy spectrum. This rotation 
causes an angular offset between the linear polarization emitted at the back of the jet
and the front, hence, generating significant levels of CP.  
In a pure pair-plasma, conversion 
of LP to CP can also occur, in the absence of internal Faraday rotation, due to a 
systematic geometrical rotation of the magnetic 
field along the line of sight through the jet \citep{hodge1982,ensslin2003}. 
One possible realisation of this is a helical magnetic field geometry, as is expected 
from jet formation models \citep[e.g.][]{mckinneyblandford2009}, but any other magnetic field 
structure which varies systematically in orientation across the jet is possible. 
We note that the sign of Stokes V for PKS~B2126--158 is the same as observed $\sim$15 years ago \citep{raynerphd}. 
Although we cannot say that it has not reversed sign an even number of times in the intervening years, 
a constant CP sign would be consistent with previous observations of other 
sources which have no evidence for a change in the sign of Stokes $V$ over decades \citep{homancpstability}. 
A preferred handedness of the CP (sign of Stokes $V$) is a strong indication 
that there is a significant component of vector-ordered poloidal field along the jet axis. 
Hence, CP generated from LP in a jet with a helical magnetic field, as described in detail by \cite{gabuzdacp2008}, 
is very plausible. In such a model, the CP sign can be used to infer the magnetic polarity of 
the jet which in turn can indicate the direction of the accretion flow and/or black-hole spin \citep{ensslin2003, gabuzdacp2008}. 

In order to derive some quantitative numbers, we begin with the work of \cite{kennettmelrose1998}, 
who defined the term relativistic rotation measure (RRM) to 
characterise the relativistic particles and magnetic field within the jet that generates  
CP through Faraday conversion. Considering a jet in approximate 
equipartition, they find ${\rm RRM}\sim10^9 B^4 L$~rad~m$^{-3}$, where $L$ is the path length 
through the jet. In the case of a purely relativistic plasma, 
of either electron-proton or electron-position composition, we expect 
$V\propto \sin({c^3{\rm RRM}/\nu^3})$ in the optically-thin regime \citep{macquart2000}. 
Fitting this relation to our data from 8 to 10 GHz, we find ${\rm RRM}=661\pm127$~rad~m$^{-3}$, 
which allows us to estimate a jet magnetic field strength $\sim$30~mG assuming 
a jet width of 1~pc. We consider this magnetic field strength reasonably robust since 
changing the jet width by an order of magnitude only changes the 
magnetic field estimate by a factor of $\sim1.8$. 

Using standard synchrotron theory and assuming that there is a single 
jet feature dominating the emission, we can estimate the angular 
size of the emission region using the equations listed in \cite{marscher1987}. 
At unity optical depth, the magnetic field strength can be estimated using
\begin{equation}
B=10^{-5} b(\alpha) S_m^{-2} \nu_m^5 \theta^4 (1+z)^{-1} \delta\,\,\,[{\rm G}]
\end{equation}
where $S_m=1.744$~Jy is the measured flux at the turnover frequency of $\nu_m=5.7$~GHz, 
$\theta$ is the angular size of the emission region in milli-arcseconds, $b(\alpha)\sim3.5$ \citep{marscher1983}, 
$z$ is the redshift and $\delta$ is the Doppler boosting factor. Using $B\sim30$~mG and 
$\delta=14$ \citep{ghisellini2010}, we find $\theta\sim0.6$~mas. Importantly, this is 
consistent with our assumption that the emission is dominated by a single, compact 
component near the base of the jet. Furthermore, we can predict that observations with 
the VLBA at 22~GHz and higher should be able to resolve this emission feature. 
The formulation of \cite{marscher1983} also allows us to independently compare the energy density 
in the relativistic electrons ($U_{re}$) with the energy density in the magnetic field ($U_B$). 
We find $U_{re}\sim3U_B$, showing that the equipartition magnetic field estimated 
through the RRM formulation is roughly consistent with the well-tested synchrotron theory. 
Intrinsic CP from synchrotron emission ($m_c^{int}$) is likely to contribute a small amount 
to the total observed CP level for a magnetic field strength of $\sim$30~mG. 
However, it is interesting to note that even if the jet magnetic field was completely uniform, which is highly 
unlikely given the low observed levels of LP, then the intrinsic 
CP generation mechanism would still not be able to reproduce the 
observed CP of $\sim$1\% at 5.7~GHz ($m_c^{int} \sim \gamma^{-1}\sim0.7\%$, 
where $\gamma= \left( \frac{\nu(1+z)/\delta}{2.8B} \right)^{1/2}$ is the Lorentz factor 
of the radiating electrons). 

In order to test the Faraday rotation-driven and magnetic-twist conversion models, 
we use the relevant equations \citep[e.g.,][and references therein]{wardle1998, ruszkowski2002, wardlehoman2003, homan2009} describing the relationship between the linear and circular polarisations 
in a partially ordered field. For small optical depths, we have
\begin{equation}
m_c\sim\frac{1}{6}p^2 \tau_F \tau_C, 
\end{equation}
where $\tau_F$ is the 
Faraday rotation depth and $\tau_C$ is the Faraday conversion depth of the jet, which 
are described by the equations (ignoring some factors of order unity)
\begin{equation}
\tau_F \sim f_u \gamma^2 \tau \frac{\ln \gamma_{\rm min}}{\gamma_{\rm min}^3},  
\end{equation} 
and
\begin{equation}
\tau_C \sim \tau \, \ln \frac{\gamma}{\gamma_{\rm min}} 
\end{equation}

\noindent where $\gamma_{\rm min}$ is the lower cutoff in the relativistic particle energy spectrum 
and $f_u$ is the fraction of uniform magnetic field along the jet axis. 
To generate $m_c\sim1\%$ from $p\lesssim0.2\%$ through Faraday conversion requires $|\tau_F|>1000$~rad~m$^{-2}$ 
at unity optical depth ($\tau=1$), where we have $\nu\sim5.7$~GHz, $B\sim30$~mG and $\delta\sim14$. 
Such a large Faraday depth model is inconsistent with our observations since we find $\tau_F=-6\pm2$~(using the observed RM of $-239\pm87$~rad~m$^{-2}$ in the frequency range from 4.5 to 6.5 GHz). 
We note that the simple scalings listed above do not hold in all cases, therefore, 
more detailed simulations are required before discarding such a 
model. 
For example, significant levels of CP can occur from anisotropic turbulence in the 
jet, where the anisotropy is created from a net poloidal magnetic flux, 
and conversion of LP to CP can occur over very small length scales in the jet 
due to very large internal Faraday rotation \citep{ruszkowski2002,beckertfalcke2002}. 
In this model, significant depolarisation of the linear polarization can occur, 
resulting in the CP exceeding the LP in some cases. 
For example, \cite{ruszkowski2002} were still able to produce a CP excess of $\sim$5 
in their model for Sgr A$^*$ with $\gamma_{\rm min}=3$ and no addition of cold 
electrons to the jet. 

A second scenario worth considering is where some of the Faraday rotation 
is external to the jet and variations in the RM ($\sigma_{\rm RM}$) about the 
mean observed value cause depolarization of the linearly polarized emission. 
Assuming random fluctuations in the magnetic field and/or electron density 
external to the jet leads to depolarisation described by 
$p/p_i = e^{-2\sigma_{\rm RM}^2 \lambda^4}$ \citep{burn1966},
where $p_i$ is the intrinsic degree of linear polarization in the jet. 
Using $p_i\sim3\%$, which we consider typical of a synchrotron self-absorbed 
VLBI core \citep{listerhoman2005}, we require $\sigma_{\rm RM}\sim400$~rad~m$^{-2}$ 
to observe $p\lesssim0.2\%$ at 5.7~GHz. While such a seemingly large dispersion in RM 
may be surprising, large spatial variations in RM of order of thousands of 
rad~m$^{-2}$ are not uncommon in the innermost 
regions of AGN jets 
\citep[e.g.,][]{attridge2005, algaba2011, algaba2013}. 
We can now obtain $m_c\sim1\%$ more easily from Faraday conversion 
of the fraction linear polarisation, $p_i\sim3\%$, within the jet, with $f_u=0.1$ 
and $\gamma_{\rm min}=7$. The Faraday depth of the jet is now 
approximately a factor of 2 greater than the observed value, however, 
a small number of reversals of the line-of-sight magnetic field through 
the jet could bring this in line with the observations. 
In summary, this model remains consistent with the observations assuming the CP 
is generated through Faraday rotation-driven conversion by the low energy 
end of the relativistic electron energy spectrum and some depolarisation of the LP 
also occurs external to the jet. 

The third main case to consider is one in which the internal Faraday rotation is 
completely suppressed due to the jet being composed of equal numbers of electrons and positrons \citep{ensslin2003}. 
The observed CP must then be generated through conversion from LP due a systematic twist 
in the jet magnetic field (e.g., a helical magnetic field), with the process being mathematically 
identical to Faraday rotation-driven conversion at a single frequency \citep[cf][]{ensslin2003}. 
Such a model is difficult to constrain since all the observed Faraday rotation 
(and frequency-dependent depolarisation of the LP) must be external to the jet. 
For PKS~B2126-158, we found the observed RM scales with frequency in a similar manner to 
what is expected for Faraday rotation internal to the jet (Section 3.2). Therefore, in this case 
the external RM would most likely come from a mixing layer/sheath surrounding the jet 
where the product $n_eB$ scales in the same manner with distance from the base of the jet 
as expected from internal Faraday rotation. 
Overall, the frequency dependent properties of the CP is one of the main aspects in which this model 
can differ observationally from the Faraday rotation-driven 
conversion case. For example, in the case of a helical 
magnetic field the magnitude of the CP depends only on $\tau_C$, and is therefore expected 
to decrease more slowly with increasing frequency than in the Faraday rotation-driven case, which 
depends on the product $\tau_F \tau_C$ \citep{ensslin2003, homan2009}.

The observed steep frequency dependence of $m_c$ above the turnover 
in Stokes $I$ (from 8 to 10~GHz) 
means that there are little or no thermal electrons mixed in with the 
relativistic plasma in the jet, as already discussed. 
Additional observational constraints and detailed jet simulations are required 
to determine the fraction of relativistic electrons to positrons and protons in the jet. 
Below the turnover (from 1.5 to 6.5~GHz), we attempt a qualitative description of 
the frequency dependence of $m_c$ using a simple conical jet model \citep{blandfordkonigl1979, wardlehoman2003}. 
Considering conservation of magnetic flux in such a model, the uniform component of the 
magnetic field along the jet axis ($B_u$) will decay as $r^{-2}$, while the fluctuating component 
of the field ($B_{rms}$) will decay as $r^{-1}$ for a conical jet in equipartition \citep[e.g.,][]{ruszkowski2002}. 
Using Eqn.~3, we see that $\tau_F\propto \gamma^2 B_u/B_{rms}$ and $\tau_C\propto \ln \gamma$ 
for relativistic electrons radiating at a particular frequency, $\nu \propto \gamma^2 B_{rms}$. 
For a constant or slowly changing $\gamma$ along the jet, $\tau_C\sim {\rm constant}$, 
while $m_c \propto \tau_F \propto \nu B_u/B_{rms}^2 \propto \nu$. Therefore, in the optically thick 
regime, the degree of CP will increase with increasing frequency as observed from 1.5 to 6.5~GHz. 
Our fitted result of $m_c\propto \nu^{+0.60\pm0.03}$ could be obtained for slight 
variations in these scalings. For example, if the fluctuating component of the magnetic 
field falls off slightly faster with distance from the base of the jet (e.g., $m_c\propto \nu^{+0.6}$ 
for $B_{rms} \propto r^{-1.2}$).

Recently, \cite{homan2009} studied the full radio polarization spectra of the quasar 
3C\,279 ($z=0.5362$) at high spatial resolution with the Very Long Baseline Array (VLBA). 
They used numerical radiative transfer simulations to test various jet models for production of 
CP that also remained consistent with the LP and total intensity emission. 
They found the Stokes $I$ core flux of 3C\,279 ($z=0.5362$) had an inverted spectrum with $I\propto\nu^{+0.9}$ 
from 8 to 24~GHz (similar to PKS\,B2126--158 from 1 to 3~GHz) while the fractional CP reached 
$\sim$1\% at 24~GHz with $p\sim2$ to 3\%. 
Through an extensive parameter-space search, 
they found a model consistent with all the observed data that implies 
the CP is most likely generated from a combination of intrinsic CP and conversion from 
Faraday rotation internal to the jet. 
From figure 10 of \cite{homan2009} we see that the expected CP varies from $\sim0.5\%$ 
at 8~GHz to $\sim1\%$ at 24~GHz leading to a CP spectral index of $\alpha_{m_c}\sim0.6$. 
This is consistent with our measured value of $\alpha_{m_c}=0.60\pm0.03$ from 
1.5 to 6.5~GHz (or $\sim$6 to 28~GHz in the rest frame of PKS\,B2126--158). 
The application of similar radiative transfer simulations to the jet emission properties 
of PKS\,2126--158 is required to determine if the CP generation mechanism is 
indeed similar to that of 3C\,279. We also need to understand why such large fractional CP 
is generated in these sources compared to the rest of the radio-loud AGN population, 
which typically have $m_c\lesssim0.3\%$ \citep{homanlister2006}.

\section{Conclusions}

From broadband radio spectropolarimetric observations with the Australia Telescope 
Compact Array (ATCA), we have obtained full-Stokes measurements of the 
quasar, PKS\,B2126--158, from 1 to 10~GHz. 
We find that the Stokes $I$ spectrum has a peak flux of $I_{max}\sim1.7$~Jy 
at a turnover frequency $\nu_t\sim5.7$~GHz with an inverted spectral index 
of $\alpha=+0.945\pm0.001$ below the turnover and a steep spectral index 
$\alpha=-0.708\pm0.003$ above the turnover. 
Left-circularly polarized emission is detected at high significance across the entire band. 
The measured Stokes $V$ has a consistent sign and varies smoothly across the 
band, with a maximum Stokes $V\sim-18$~mJy at 6.5~GHz ($m_c\sim1\%$), 
before the spectrum turns over somewhere between 6.5 and 8~GHz. 
Our detection exhibits the same sign as found 15 years ago by \cite{raynerphd}, 
who obtained $V/I\sim-1.4\%$ at 4.8~GHz. 

From comparison of the integrated flux with high spatial resolution VLBI images, 
we conclude that there is no significant amount of flux on intermediate angular scales 
and that effectively all the emission comes from the compact inner jet regions 
(on scales $<5$~mas). Hence, our fractional linear polarization ($p$) and circular 
polarization (CP) data should be 
equivalent to that obtained from VLBI images at the same frequencies. 
The LP is clearly anti-correlated with the CP, with the degree of CP 
$m_c\propto p^{-0.24\pm0.03}$, where $m_c\equiv |V|/I$. 
This is the first time such a relation has been observed and clearly indicates the 
action of Faraday conversion of LP to CP within the jet. 

By fitting a power-law to the frequency variation in the degree of CP
above the turnover, we find a very steep CP spectral dependence, 
$m_c\propto \nu^{-3.0\pm0.4}$, which is consistent with the prediction of 
$m_c\propto \nu^{-3}$ for conversion of linear polarization (LP) to CP by 
purely relativistic particles and magnetic fields within the jet (i.e., no thermal 
plasma within the jet). 
Below the turnover we find $m_c\propto \nu^{+0.60\pm0.03}$. The increase of $m_c$ 
with frequency in the optically thick regime is easily understood 
considering the frequency-dependent location of the emission in which higher 
frequency observations probe further upstream in the jet where the uniform 
component of the magnetic field is stronger and, hence, the amount of $m_c$ 
due to both Faraday conversion and intrinsic CP is larger. 
 
Overall, our results conclusively favour Faraday conversion of LP to CP 
within the jet as the dominant CP generation mechanism in this source. 
We are unable to uniquely constrain whether the conversion is achieved 
by Faraday rotation within the jet dominated by the low energy end of the 
relativistic electron energy spectrum, or by a change in the orientation of the 
perpendicular component of a vector-ordered magnetic field through the jet 
(e.g. a helical magnetic field). 
Three different scenarios are discussed: \\1. Very large amounts of internal Faraday 
rotation could naturally produce the large CP-to-LP ratios of 0.5 to 10 across our 
observed frequency range, but this appears inconsistent with our RM measurements. 
\\2. Considering our measured RM as being composed of some internal and 
external components, we can obtain the observed levels of CP at $\tau\sim1$ 
due to Faraday rotation-driven conversion, with the fraction of uniform magnetic 
field along the jet axis $f_u=0.1$ and a low energy cut-off in the relativistic 
electron energy spectrum of $\gamma_{min}=7$. 
\\3. If all the observed Faraday rotation is external to the jet, then we require 
large spatial gradients in the external RM distribution to strongly depolarise 
the LP emission. Conversion of LP to CP within the jet is then achieved solely by 
a systematic twist in the magnetic field through an electron-positron jet.

The observed low levels of LP and frequency dependence of the magnitude of 
the RM favour a model in which there is some contribution of internal Faraday 
rotation. This leads us to suggest that Faraday rotation-driven conversion by 
the low energy end of the relativistic electron energy spectrum, in a mainly 
electron-proton jet with $\sim$10\% uniform magnetic flux along the jet axis, 
as the most likely explanation for the observed levels of CP in this source. 
Future work requires detailed numerical, polarized, radiative transfer simulations in 
conjunction with realistic internal and external Faraday 
rotation models to consistently explain the full-Stokes 
emission from PKS\,B2126--158 and its variation 
across the entire observed frequency range.

\section{Acknowledgements}

The Australia Telescope Compact Array is part of the Australia Telescope, which 
is funded by the Commonwealth of Australia for operation as a National Facility 
managed by CSIRO.
S.O'S and B.M.G. acknowledge the 
support of the Australian Research Council through grants FS100100033 
and FL100100114, respectively. 
The authors wish to thank J.~P.~Macquart for useful discussions as well as 
M.~A.~Lara-L\'opez for help with the figures. 
We also thank the referee, John Wardle, for helpful comments and discussion 
which significantly improved this paper. 
This research has made use of NASA's Astrophysics Data System Service and 
the NASA/IPAC Extragalactic Database (NED) which is operated by the Jet 
Propulsion Laboratory, California Institute of Technology, under contract with the 
National Aeronautics and Space Administration. 
This research has also made use of the SIMBAD database,
operated at CDS, Strasbourg, France.

  \bibliographystyle{mn2e}
  \bibliography{cp_bib}

\begin{thebibliography}{}

\bibitem[\protect\citeauthoryear{{Algaba}}{{Algaba}}{2013}]{algaba2013}
{Algaba} J.~C.,  2013, MNRAS, in press, arXiv:1212.3423

\bibitem[\protect\citeauthoryear{{Algaba}, {Gabuzda} \& {Smith}}{{Algaba}
  et~al.}{2011}]{algaba2011}
{Algaba} J.~C.,  {Gabuzda} D.~C.,    {Smith} P.~S.,  2011, MNRAS, 411, 85

\bibitem[\protect\citeauthoryear{{Aller} \& {Aller}}{{Aller} \&
  {Aller}}{2012}]{aller2012}
{Aller} H.~D.,  {Aller} M.~F.,  2012, BAAS, 220, 335.20

\bibitem[\protect\citeauthoryear{Attridge, Wardle \& Homan}{Attridge
  et~al.}{2005}]{attridge2005}
Attridge J.~M.,  Wardle J. F.~C.,    Homan D.~C.,  2005, ApJ, 633, L85

\bibitem[\protect\citeauthoryear{Beckert \& Falcke}{Beckert \&
  Falcke}{2002}]{beckertfalcke2002}
Beckert T.,  Falcke H.,  2002, A\&A, 388, 1106

\bibitem[\protect\citeauthoryear{Begelman, Blandford \& Rees}{Begelman
  et~al.}{1984}]{begelmanblandfordrees1984}
Begelman M.~C.,  Blandford R.~D.,    Rees M.~J.,  1984, Reviews of Modern
  Physics, 56, 255

\bibitem[\protect\citeauthoryear{{Bicknell}, {Dopita} \& {O'Dea}}{{Bicknell}
  et~al.}{1997}]{bicknell1997}
{Bicknell} G.~V.,  {Dopita} M.~A.,    {O'Dea} C.~P.~O.,  1997, ApJ, 485, 112

\bibitem[\protect\citeauthoryear{Blandford \& K\"onigl}{Blandford \&
  K\"onigl}{1979}]{blandfordkonigl1979}
Blandford R.~D.,  K\"onigl A.,  1979, ApJ, 232, 34

\bibitem[\protect\citeauthoryear{{Blandford} \& {Payne}}{{Blandford} \&
  {Payne}}{1982}]{blandfordpayne1982}
{Blandford} R.~D.,  {Payne} D.~G.,  1982, MNRAS, 199, 883

\bibitem[\protect\citeauthoryear{{Blandford} \& {Znajek}}{{Blandford} \&
  {Znajek}}{1977}]{blandfordznajek1977}
{Blandford} R.~D.,  {Znajek} R.~L.,  1977, MNRAS, 179, 433

\bibitem[\protect\citeauthoryear{{Broderick} \& {Loeb}}{{Broderick} \&
  {Loeb}}{2009}]{broderickloeb2009}
{Broderick} A.~E.,  {Loeb} A.,  2009, ApJ, 703, L104

\bibitem[\protect\citeauthoryear{{Burn}}{{Burn}}{1966}]{burn1966}
{Burn} B.~J.,  1966, MNRAS, 133, 67

\bibitem[\protect\citeauthoryear{{Celotti}, {Kuncic}, {Rees} \&
  {Wardle}}{{Celotti} et~al.}{1998}]{celotti1998}
{Celotti} A.,  {Kuncic} Z.,  {Rees} M.~J.,    {Wardle} J.~F.~C.,  1998, MNRAS,
  293, 288

\bibitem[\protect\citeauthoryear{{Charlot}, {Boboltz}, {Fey}, {Fomalont},
  {Geldzahler}, {Gordon}, {Jacobs}, {Lanyi} \& {Ma}}{{Charlot}
  et~al.}{2010}]{charlot}
{Charlot} P.,  {Boboltz} D.~A.,  {Fey} A.~L.,  {Fomalont} E.~B.,  {Geldzahler}
  B.~J.,  {Gordon} D.,  {Jacobs} C.~S.,  {Lanyi} G.~E.,    {Ma} C.,  2010, AJ,
  139, 1713

\bibitem[\protect\citeauthoryear{{D'Odorico}, {Cristiani}, {D'Odorico},
  {Fontana} \& {Giallongo}}{{D'Odorico} et~al.}{1998}]{dodorico1998}
{D'Odorico} V.,  {Cristiani} S.,  {D'Odorico} S.,  {Fontana} A.,    {Giallongo}
  E.,  1998, A\&AS, 127, 217

\bibitem[\protect\citeauthoryear{{En{\ss}lin}}{{En{\ss}lin}}{2003}]{ensslin2003}
{En{\ss}lin} T.~A.,  2003, A\&A, 401, 499

\bibitem[\protect\citeauthoryear{{Farnsworth}, {Rudnick} \&
  {Brown}}{{Farnsworth} et~al.}{2011}]{farnsworth2011}
{Farnsworth} D.,  {Rudnick} L.,    {Brown} S.,  2011, AJ, 141, 191

\bibitem[\protect\citeauthoryear{{Fomalont}, {Frey}, {Paragi}, {Gurvits},
  {Scott}, {Taylor}, {Edwards} \& {Hirabayashi}}{{Fomalont}
  et~al.}{2000}]{fomalont}
{Fomalont} E.~B.,  {Frey} S.,  {Paragi} Z.,  {Gurvits} L.~I.,  {Scott} W.~K.,
  {Taylor} A.~R.,  {Edwards} P.~G.,    {Hirabayashi} H.,  2000, ApJS, 131, 95

\bibitem[\protect\citeauthoryear{Gabuzda, Vitrishchak, Mahmud \&
  O'Sullivan}{Gabuzda et~al.}{2008}]{gabuzdacp2008}
Gabuzda D.~C.,  Vitrishchak V.~M.,  Mahmud M.,    O'Sullivan S.~P.,  2008,
  MNRAS, 384, 1003

\bibitem[\protect\citeauthoryear{{Ghisellini}, {Della Ceca}, {Volonteri},
  {Ghirlanda}, {Tavecchio}, {Foschini}, {Tagliaferri} \& {Haardt}}{{Ghisellini}
  et~al.}{2010}]{ghisellini2010}
{Ghisellini} G.,  {Della Ceca} R.,  {Volonteri} M.,  {Ghirlanda} G.,
  {Tavecchio} F.,  {Foschini} L.,  {Tagliaferri} G.,    {Haardt} F.,  2010,
  MNRAS, 405, 387

\bibitem[\protect\citeauthoryear{{Hodge}}{{Hodge}}{1982}]{hodge1982}
{Hodge} P.~E.,  1982, ApJ, 263, 595

\bibitem[\protect\citeauthoryear{{Homan}, {Attridge} \& {Wardle}}{{Homan}
  et~al.}{2001}]{homan2001}
{Homan} D.~C.,  {Attridge} J.~M.,    {Wardle} J.~F.~C.,  2001, ApJ, 556, 113

\bibitem[\protect\citeauthoryear{{Homan} \& {Lister}}{{Homan} \&
  {Lister}}{2006}]{homanlister2006}
{Homan} D.~C.,  {Lister} M.~L.,  2006, AJ, 131, 1262

\bibitem[\protect\citeauthoryear{{Homan}, {Lister}, {Aller}, {Aller} \&
  {Wardle}}{{Homan} et~al.}{2009}]{homan2009}
{Homan} D.~C.,  {Lister} M.~L.,  {Aller} H.~D.,  {Aller} M.~F.,    {Wardle}
  J.~F.~C.,  2009, ApJ, 696, 328

\bibitem[\protect\citeauthoryear{{Homan}, {Lister} \& {MOJAVE}}{{Homan}
  et~al.}{2011}]{homancpstability}
{Homan} D.~C.,  {Lister} M.~L.,    {MOJAVE} 2011, BAAS, 43, 310.03

\bibitem[\protect\citeauthoryear{{Homan} \& {Wardle}}{{Homan} \&
  {Wardle}}{1999}]{homanwardle1999}
{Homan} D.~C.,  {Wardle} J.~F.~C.,  1999, AJ, 118, 1942

\bibitem[\protect\citeauthoryear{{Homan} \& {Wardle}}{{Homan} \&
  {Wardle}}{2004}]{homanwardle2004}
{Homan} D.~C.,  {Wardle} J.~F.~C.,  2004, ApJL, 602, L13

\bibitem[\protect\citeauthoryear{{Jauncey}, {Wright}, {Peterson} \&
  {Condon}}{{Jauncey} et~al.}{1978}]{jauncey1978}
{Jauncey} D.~L.,  {Wright} A.~E.,  {Peterson} B.~A.,    {Condon} J.~J.,  1978,
  ApJL, 223, L1

\bibitem[\protect\citeauthoryear{{Jones}}{{Jones}}{1988}]{jones1988}
{Jones} T.~W.,  1988, ApJ, 332, 678

\bibitem[\protect\citeauthoryear{{Jones} \& {O'Dell}}{{Jones} \&
  {O'Dell}}{1977}]{jonesodell1977b}
{Jones} T.~W.,  {O'Dell} S.~L.,  1977, ApJ, 215, 236

\bibitem[\protect\citeauthoryear{Jorstad, Marscher, Lister, Stirling,
  Cawthorne, Gear, G\'omez, Stevens, Smith, Forster \& Robson}{Jorstad
  et~al.}{2005}]{jorstad2005}
Jorstad S.~G.,  Marscher A.~P.,  Lister M.~L.,  Stirling A.~M.,  Cawthorne
  T.~V.,  Gear W.~K.,  G\'omez J.~L.,  Stevens J.~A.,  Smith P.~S.,  Forster
  J.~R.,    Robson E.~I.,  2005, AJ, 130, 1418

\bibitem[\protect\citeauthoryear{{Kennett} \& {Melrose}}{{Kennett} \&
  {Melrose}}{1998}]{kennettmelrose1998}
{Kennett} M.,  {Melrose} D.,  1998, PASA, 15, 211

\bibitem[\protect\citeauthoryear{{Komesaroff}, {Roberts}, {Milne}, {Rayner} \&
  {Cooke}}{{Komesaroff} et~al.}{1984}]{komesaroff1984}
{Komesaroff} M.~M.,  {Roberts} J.~A.,  {Milne} D.~K.,  {Rayner} P.~T.,
  {Cooke} D.~J.,  1984, MNRAS, 208, 409

\bibitem[\protect\citeauthoryear{{Legg} \& {Westfold}}{{Legg} \&
  {Westfold}}{1968}]{leggwestfold1968}
{Legg} M.~P.~C.,  {Westfold} K.~C.,  1968, ApJ, 154, 499

\bibitem[\protect\citeauthoryear{{Lister} \& {Homan}}{{Lister} \&
  {Homan}}{2005}]{listerhoman2005}
{Lister} M.~L.,  {Homan} D.~C.,  2005, AJ, 130, 1389

\bibitem[\protect\citeauthoryear{{Macquart}}{{Macquart}}{2002}]{macquart2002}
{Macquart} J.-P.,  2002, PASA, 19, 43

\bibitem[\protect\citeauthoryear{{Macquart} \& {Fender}}{{Macquart} \&
  {Fender}}{2004}]{macquartfender2004}
{Macquart} J.-P.,  {Fender} R.~P.,  eds, 2004, {Circular Polarisation from
  Relativistic Jet Sources}.
Kluwer Academic Publishers, Dordrecht.

\bibitem[\protect\citeauthoryear{{Macquart}, {Kedziora-Chudczer}, {Rayner} \&
  {Jauncey}}{{Macquart} et~al.}{2000}]{macquart2000}
{Macquart} J.-P.,  {Kedziora-Chudczer} L.,  {Rayner} D.~P.,    {Jauncey} D.~L.,
   2000, ApJ, 538, 623

\bibitem[\protect\citeauthoryear{{Macquart} \& {Melrose}}{{Macquart} \&
  {Melrose}}{2000}]{macquartmelrose2000}
{Macquart} J.-P.,  {Melrose} D.~B.,  2000, ApJ, 545, 798

\bibitem[\protect\citeauthoryear{{Marscher}}{{Marscher}}{1983}]{marscher1983}
{Marscher} A.~P.,  1983, ApJ, 264, 296

\bibitem[\protect\citeauthoryear{Marscher}{Marscher}{1987}]{marscher1987}
Marscher A.~P.,  1987, in Zensus J.~A.,  Pearson T.~J.,  eds, Superluminal
  Radio Sources. Vol.~280.
Cambridge Univ. Press

\bibitem[\protect\citeauthoryear{{McKinney} \& {Blandford}}{{McKinney} \&
  {Blandford}}{2009}]{mckinneyblandford2009}
{McKinney} J.~C.,  {Blandford} R.~D.,  2009, MNRAS, 394, L126

\bibitem[\protect\citeauthoryear{Meier}{Meier}{2009}]{meierjapan}
Meier D.~L.,  2009, in Hagiwara Y.,  Fomalont E.,  Tsuboi M.,   Murata Y.,
  eds, Approaching Micro-Arcsecond Resolution with VSOP-2: Astrophysics and
  Technology. Vol.~402.
ASP Conf. Ser., San Francisco, p.~342

\bibitem[\protect\citeauthoryear{{Melrose}}{{Melrose}}{1972}]{melrose1972}
{Melrose} D.~B.,  1972, PASA, 2, 140

\bibitem[\protect\citeauthoryear{{Mutel}, {Hodges} \& {Phillips}}{{Mutel}
  et~al.}{1985}]{mutel1985}
{Mutel} R.~L.,  {Hodges} M.~W.,    {Phillips} R.~B.,  1985, ApJ, 290, 86

\bibitem[\protect\citeauthoryear{{Osmer}, {Porter} \& {Green}}{{Osmer}
  et~al.}{1994}]{osmer1994}
{Osmer} P.~S.,  {Porter} A.~C.,    {Green} R.~F.,  1994, ApJ, 436, 678

\bibitem[\protect\citeauthoryear{{O'Sullivan}, {Brown}, {Robishaw},
  {Schnitzeler}, {McClure-Griffiths}, {Feain}, {Taylor}, {Gaensler},
  {Landecker}, {Harvey-Smith} \& {Carretti}}{{O'Sullivan}
  et~al.}{2012}]{osullivan2012}
{O'Sullivan} S.~P.,  {Brown} S.,  {Robishaw} T.,  {Schnitzeler} D.~H.~F.~M.,
  {McClure-Griffiths} N.~M.,  {Feain} I.~J.,  {Taylor} A.~R.,  {Gaensler}
  B.~M.,  {Landecker} T.~L.,  {Harvey-Smith} L.,    {Carretti} E.,  2012,
  MNRAS, 421, 3300

\bibitem[\protect\citeauthoryear{O'Sullivan \& Gabuzda}{O'Sullivan \&
  Gabuzda}{2009}]{osullivangabuzda2009}
O'Sullivan S.~P.,  Gabuzda D.~C.,  2009, MNRAS, 393, 429

\bibitem[\protect\citeauthoryear{Pacholczyk}{Pacholczyk}{1973}]{pacho1973}
Pacholczyk A.~G.,  1973, MNRAS, 163, 29

\bibitem[\protect\citeauthoryear{{Porth}, {Fendt}, {Meliani} \&
  {Vaidya}}{{Porth} et~al.}{2011}]{porth2011}
{Porth} O.,  {Fendt} C.,  {Meliani} Z.,    {Vaidya} B.,  2011, ApJ, 737, 42

\bibitem[\protect\citeauthoryear{{Rayner}}{{Rayner}}{2000}]{raynerphd}
{Rayner} D.~P.,  2000, PASA, 17, 284

\bibitem[\protect\citeauthoryear{{Rayner}, {Norris} \& {Sault}}{{Rayner}
  et~al.}{2000}]{rayner2000}
{Rayner} D.~P.,  {Norris} R.~P.,    {Sault} R.~J.,  2000, MNRAS, 319, 484

\bibitem[\protect\citeauthoryear{{Ruszkowski} \& {Begelman}}{{Ruszkowski} \&
  {Begelman}}{2002}]{ruszkowski2002}
{Ruszkowski} M.,  {Begelman} M.~C.,  2002, ApJ, 573, 485

\bibitem[\protect\citeauthoryear{{Saikia} \& {Salter}}{{Saikia} \&
  {Salter}}{1988}]{saikiasalter1988}
{Saikia} D.~J.,  {Salter} C.~J.,  1988, ARA\&A, 26, 93

\bibitem[\protect\citeauthoryear{{Sault}, {Teuben} \& {Wright}}{{Sault}
  et~al.}{1995}]{sault1995}
{Sault} R.~J.,  {Teuben} P.~J.,    {Wright} M.~C.~H.,  1995, in {R.~A.~Shaw,
  H.~E.~Payne, \& J.~J.~E.~Hayes} ed., Astronomical Data Analysis Software and
  Systems IV Vol.~77 of ASP Conference Series.
p.~433

\bibitem[\protect\citeauthoryear{{Scott}, {Fomalont}, {Horiuchi}, {Lovell},
  {Moellenbrock}, {Dodson}, {Edwards}, {Coldwell} \& {Fodor}}{{Scott}
  et~al.}{2004}]{scott}
{Scott} W.~K.,  {Fomalont} E.~B.,  {Horiuchi} S.,  {Lovell} J.~E.~J.,
  {Moellenbrock} G.~A.,  {Dodson} R.~G.,  {Edwards} P.~G.,  {Coldwell} G.~V.,
   {Fodor} S.,  2004, ApJS, 155, 33

\bibitem[\protect\citeauthoryear{{Stanghellini}}{{Stanghellini}}{2003}]{stanghellini2003}
{Stanghellini} C.,  2003, PASA, 20, 118

\bibitem[\protect\citeauthoryear{{Stanghellini}, {O'Dea}, {Dallacasa}, {Baum},
  {Fanti} \& {Fanti}}{{Stanghellini} et~al.}{1998}]{stanghellini1998}
{Stanghellini} C.,  {O'Dea} C.~P.,  {Dallacasa} D.,  {Baum} S.~A.,  {Fanti} R.,
     {Fanti} C.,  1998, A\&AS, 131, 303

\bibitem[\protect\citeauthoryear{{Taylor}, {Stil} \& {Sunstrum}}{{Taylor}
  et~al.}{2009}]{taylor2009}
{Taylor} A.~R.,  {Stil} J.~M.,    {Sunstrum} C.,  2009, ApJ, 702, 1230

\bibitem[\protect\citeauthoryear{{Tingay}, {Jauncey}, {King}, {Tzioumis},
  {Lovell} \& {Edwards}}{{Tingay} et~al.}{2003}]{tingay}
{Tingay} S.~J.,  {Jauncey} D.~L.,  {King} E.~A.,  {Tzioumis} A.~K.,  {Lovell}
  J.~E.~J.,    {Edwards} P.~G.,  2003, PASJ, 55, 351

\bibitem[\protect\citeauthoryear{{Tzioumis}, {King}, {Morganti}, {Dallacasa},
  {Tadhunter}, {Fanti}, {Reynolds}, {Jauncey} \& {Preston}}{{Tzioumis}
  et~al.}{2002}]{tzioumis2002}
{Tzioumis} A.,  {King} E.,  {Morganti} R.,  {Dallacasa} D.,  {Tadhunter} C.,
  {Fanti} C.,  {Reynolds} J.,  {Jauncey} D.,    {Preston} R.,  2002, A\&A, 392,
  841

\bibitem[\protect\citeauthoryear{{Valtaoja}}{{Valtaoja}}{1984}]{valtaoja1984}
{Valtaoja} E.,  1984, A\&SS, 100, 227

\bibitem[\protect\citeauthoryear{{Vitrishchak}, {Gabuzda}, {Algaba},
  {Rastorgueva}, {O'Sullivan} \& {O'Dowd}}{{Vitrishchak}
  et~al.}{2008}]{vitrishchak2008}
{Vitrishchak} V.~M.,  {Gabuzda} D.~C.,  {Algaba} J.~C.,  {Rastorgueva} E.~A.,
  {O'Sullivan} S.~P.,    {O'Dowd} A.,  2008, MNRAS, 391, 124

\bibitem[\protect\citeauthoryear{Walker, Dhawan, Romney, Kellermann \&
  Vermeulen}{Walker et~al.}{2000}]{walker2000}
Walker R.~C.,  Dhawan V.,  Romney J.~D.,  Kellermann K.~I.,    Vermeulen R.~C.,
   2000, The Astrophysical Journal, 530, 233

\bibitem[\protect\citeauthoryear{{Wardle} \& {Homan}}{{Wardle} \&
  {Homan}}{2003}]{wardlehoman2003}
{Wardle} J.~F.~C.,  {Homan} D.~C.,  2003, A\&SS, 288, 143

\bibitem[\protect\citeauthoryear{Wardle, Homan, Ojha \& Roberts}{Wardle
  et~al.}{1998}]{wardle1998}
Wardle J. F.~C.,  Homan D.~C.,  Ojha R.,    Roberts D.~H.,  1998, Nature, 395,
  457

\bibitem[\protect\citeauthoryear{{Wardle} \& {Kronberg}}{{Wardle} \&
  {Kronberg}}{1974}]{wardlekronberg1974}
{Wardle} J.~F.~C.,  {Kronberg} P.~P.,  1974, ApJ, 194, 249

\bibitem[\protect\citeauthoryear{{Weiler} \& {de Pater}}{{Weiler} \& {de
  Pater}}{1983}]{weilerdepater1983}
{Weiler} K.~W.,  {de Pater} I.,  1983, ApJS, 52, 293

\bibitem[\protect\citeauthoryear{{Wilson}, {Ferris}, {Axtens}, {Brown},
  {Davis}, {Hampson}, {Leach}, {Roberts} \& {Saunders}}{{Wilson}
  et~al.}{2011}]{cabbpaper}
{Wilson} W.~E.,  {Ferris} R.~H.,  {Axtens} P.,  {Brown} A.,  {Davis} E.,
  {Hampson} G.,  {Leach} M.,  {Roberts} P.,    {Saunders} S.,  2011, MNRAS,
  416, 832

\end{thebibliography}

\end{document}